# DEMETER: Efficient simultaneous curation of genome-scale reconstructions guided by experimental data and refined gene annotations

Almut Heinken[1,2], Stefanía Magnúsdóttir[3], Ronan M.T. Fleming[1,4], and Ines Thiele[1,2,5,6*]

[1]School of Medicine, National University of Galway, Galway, Ireland,
[2]Ryan Institute, National University of Galway, Galway, Ireland,
[3]Center for Molecular Medicine, University Medical Center Utrecht, Utrecht, The Netherlands,
[4]Leiden Academic Centre for Drug Research, Leiden University, Leiden, The Netherlands
[5]Division of Microbiology, National University of Galway, Galway, Ireland,
[6]APC Microbiome Ireland, Ireland

*To whom correspondence should be addressed.

## Abstract

**Motivation:** Manual curation of genome-scale reconstructions is laborious, yet existing automated curation tools typically do not take species-specific experimental data and manually refined genome annotations into account.

**Results:** We developed DEMETER, a COBRA Toolbox extension that enables the efficient simultaneous refinement of thousands of draft genome-scale reconstructions while ensuring adherence to the quality standards in the field, agreement with available experimental data, and refinement of pathways based on manually refined genome annotations.

**Availability:** DEMETER and tutorials are available at https://github.com/opencobra/cobratoolbox.

**Contact:** ines.thiele@nuigalway.ie

# Introduction

The Constraint-Based Reconstruction and Analysis (COBRA) approach relies on genome-scale reconstructions that have been curated based on genomic, biochemical, and physiological data, a laborious process consisting of 96 steps (Thiele and Palsson, 2010). On the other hand, existing automated reconstruction pipelines such as ModelSEED (Seaver, et al., 2021) provide limited support for curation based on organism-specific experimental and genomic data.

Here, we present DEMETER (Data-drivEn METabolic nEtwork Refinement), a reconstruction pipeline that enables the efficient and simultaneous refinement of thousands of draft genome-scale reconstructions. Previously, DEMETER enabled the reconstruction of 773 human gut microbes, AGORA (Magnusdottir, et al., 2017), as well its expansion, AGORA2, accounting for 7,206 human microbial strains (Heinken, et al., 2020). Refinement of draft reconstructions in DEMETER is guided by a wealth of experimental data, such as carbon sources, fermentation pathways, and growth requirements, for over 1,000 species, as well as by strain-specific comparative genomic analyses. Hence, DEMETER ensures the resulting refined reconstructions capture the known traits of the target organisms.

# Features

The DEMETER pipeline consists of three main steps: (i) data collection and integration, (ii) draft reconstruction refinement, testing, and debugging, and (iii) computation of model properties (Figure 1).

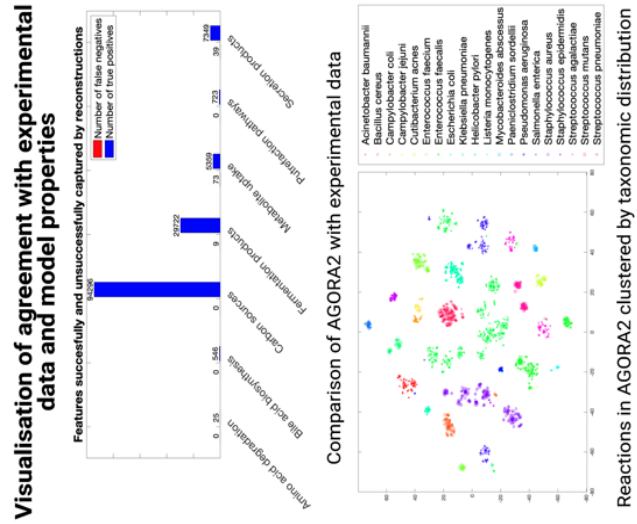

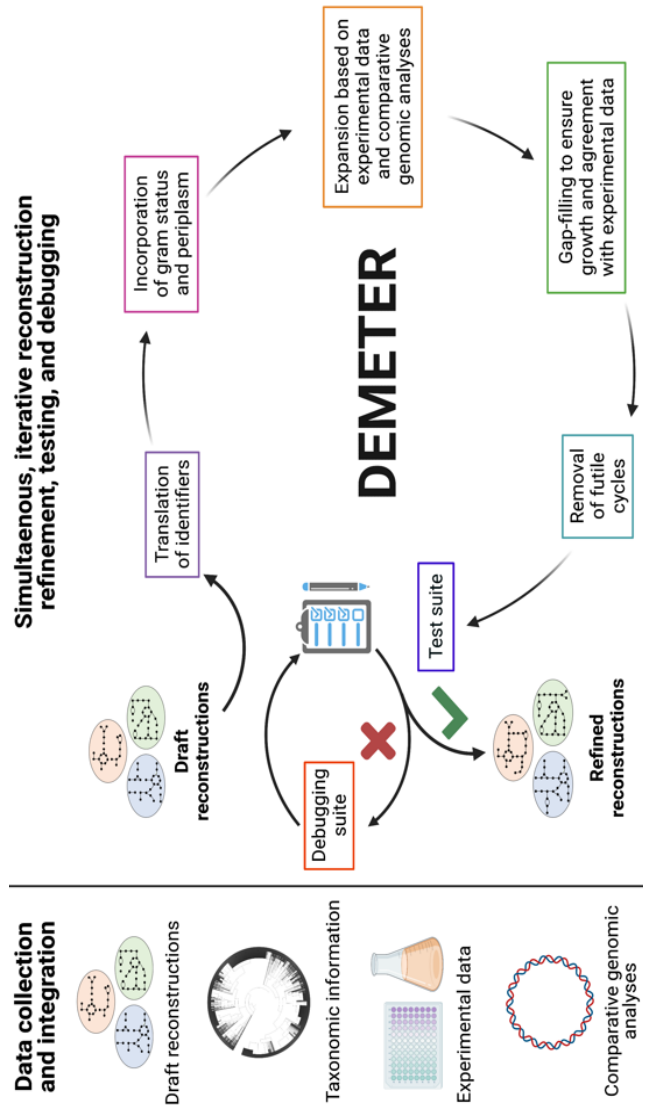

**Figure 1**: Overview of the DEMETER workflow consisting of (i) data collection and integration, (ii) simultaneous refinement, testing, and debugging of the draft reconstructions, and (iii) visualisation of test results and computation of model properties. Created with BioRender.com.

**Data collection and integration**

The minimal prerequisite the availability of a sequenced genome for the organisms of interest. An essential step is the generation of draft genome-scale reconstructions by the ModelSEED (Seaver, et al., 2021) or KBase (Arkin, et al., 2018) online platforms. Where possible, gram status and species-specific experimental data are propagated to the target organisms. Moreover, strain-specific comparative genomic analyses retrieved from PubSEED subsystems (Aziz, et al., 2012) can be mapped to DEMETER.

**Refinement step**

During the refinement step, the draft reconstructions are systematically improved (Figure 1). Briefly, the following steps are performed:

- Translation from Model SEED to Virtual Metabolic Human (Noronha, et al., 2019) reaction and metabolite nomenclature.
- Curation of the biomass objective function based on gram status and, where appropriate, generation of a periplasmatic compartment.
- Gap-filling to enable anaerobic growth on a simulated complex medium and, where available, defined media reported in the literature.
- Inclusion of species-species pathways for carbon source utilisation, fermentation products, and consumed and secreted metabolites.
- Refinement of pathways and gene-protein-reaction associations based on strain-specific comparative genomic analyses.
- Removal of futile cycles to ensure thermodynamic feasibility.
- Quality-controlled rebuilding of the resulting refined reconstruction.

**Test and debugging suite**

To ensure that refined reconstructions generated by DEMETER agree with the input data and conform to the quality standards in the field, a test suite is provided that performs systematic quality control and quality assurance (Figure 1). Any errors may subsequently be corrected through a provided automated debugging suite.

**Analysis of model properties**

To elucidate how metabolic traits are spread across strains, model features including reaction and metabolite content, metabolite uptake and secretion potential, and internal metabolite biosynthesis potential are computed and subsequently visualised. Taxonomically close strains reconstructed by DEMETER are also similar in their reaction content (Figure 1).

## Implementation

DEMETER is written in MATLAB (Mathworks, Inc.) and relies on functions implemented in the COBRA Toolbox (Heirendt, et al., 2019). For efficient large-scale processing of reconstructions, parallelisation has been implemented. A comprehensive tutorial is provided in form of a MATLAB live script (Supplementary File 1).

## Discussion

Refined reconstructions built through DEMETER adhere to the quality standards in the COBRA field and capture the known metabolic features of the target organisms. Hence, they are suitable for predictive modelling studies, such as the construction and interrogation of personalised microbiome models. Note that while DEMETER was initially developed for the human microbiome, it can be applied to any bacterial or archaeal species.

## Funding

This study was funded by grants from the European Research Council (ERC) under the European Union's Horizon 2020 research and innovation programme (grant agreement No 757922) to IT, and by the National Institute on Aging grants (1RF1AG058942-01 and 1U19AG063744-01).

*Conflict of Interest:* none declared.